\documentstyle[preprint,aps,epsfig,here]{revtex}
\begin{document}

\title{Non-singlet structure function of the $^3$He-$^3$H system and divergence of the Gottfried integral}

\author{V. Guzey$^a$, K. Saito$^b$, M. Strikman$^c$, A.W. Thomas$^a$ and K.~Tsushima$^a$}

\address{$^a$ Department of Physics and Mathematical Physics, and\\
Special Research Centre for the Subatomic Structure of Matter (CSSM),\\
Adelaide University, Adelaide SA 5005 Australia,\\
$^b$ Tohoku College of Pharmacy, Sendai 981-8558, Japan\\
$^c$ The Pennsylvania State University, University Park, PA 16802, USA} 

%\date{\today}

\preprint{
\vbox{
\hbox{ADP-01-03/T438}
}}

\maketitle

\begin{abstract}

We study shadowing and antishadowing corrections to the flavor 
non-singlet structure function $F_2^{^3{\rm He}}-F_{2}^{^3{\rm H}}$   
and show that the difference 
between the one-particle
density distributions of $^3$He and $^3$H plays an important 
role at very small $x$.  We find that the flavor non-singlet structure 
function in these mirror nuclei is enhanced at small $x$ by  nuclear 
shadowing,
which increases  the nuclear Gottfried integral, integrated from $10^{-4}$ to 1, by $11 \div 36$\%. 
When integrated from zero, the Gottfried integral is divergent for these mirror nuclei. It seems likely that, as a consequence of charge symmetry breaking, this may also apply to  the proton-neutron system.

\end{abstract}

\newpage

\section{Introduction}
\label{sec:intro}

The measurement of the flavor 
non-singlet\footnote{Obviously, the SU(3) flavor symmetry is broken by non-perturbative QCD effects, and 
the ``singlet'' and ``non-singlet'' combinations of structure functions do not transform as pure singlet and non-singlet under  SU(3) rotations. In this work, we use 
the terms ``singlet'' and ``non-singlet'' just to indicate the quark content of the corresponding structure functions.} 
 structure function $F_{2}^{p}(x,Q^2)-F_{2}^{n}(x,Q^2)$, where $F_{2}^{p}(x,Q^2)$ ($F_{2}^{n}(x,Q^2)$) is the proton (neutron) structure function, in deep inelastic muon-hydrogen and muon-deuterium scattering experiments, performed by the NMC collaboration \cite{NMC1},  led to a surprising result. The data revealed an excess of sea down quarks as compared to sea up quarks 
in the free proton. 
This conclusion has been confirmed by the E866/NuSea experiment, where the difference $\bar{d}-\bar{u}$ was measured directly using the Drell-Yan production of $\mu^{+} \mu^{-}$ pairs in proton-proton and proton-deuteron collisions \cite{E866}.

The results of both experiments contradict the expectation of perturbative QCD 
(pQCD) 
that $\bar{u} \approx \bar{d}$ in the proton. Within the framework of pQCD, the light quark sea is flavor symmetric with a good accuracy since it is generated by the perturbative splitting $g \rightarrow q \bar{q}$, which does not distinguish between the $u$ and $d$ flavors. The obvious 
inconsistency 
of the experimental data with pQCD predictions indicates that non-perturbative effects are responsible for creating flavor asymmetry in the light sea quarks.

The excess of $\bar{d}$ over $\bar{u}$ was anticipated well before the measurement on the basis of the chiral structure of QCD \cite{AWT83}. Since the NMC experimental discovery and earlier experimental indications that  $\bar{d} \neq \bar{u}$ in the proton, this explanation has been actively investigated \cite{pion}, with the latest discussion centering on the model-independent leading nonanalytic contribution \cite{TMS00}. Another possible contribution involving the Pauli principle, was first explored in pQCD, where it was found to give a negligible effect \cite{RS79}. In contrast, non-perturbative calculations based on the change in the Dirac sea in the presence of a confining potential \cite{ST89}
(for recent reviews of relevant models also see \cite{Kumano,Vogt}) as well as 
calculations \cite{soliton}, based on  
the chiral quark-soliton model of Ref.\ \cite{Diakonov},
also predict $\bar{d} > \bar{u}$, which could be of the magnitude observed experimentally.
 Both of these explanations offer considerable insight into the nature of hadronic structure in QCD and it is vital to find experimental ways to separate them.

One way to learn more about the non-perturbative dynamics of the nucleon is to consider the
non-singlet structure function $F_{2}^{p}(x,Q^2)-F_{2}^{n}(x,Q^2)$ for bound 
nucleons \cite{Bissey}. 
In this case, any discrepancy between theoretical predictions and data would indicate that the mechanisms, which explained the  $\bar{u} \neq \bar{d}$ asymmetry for the free proton, are modified in nuclear medium. The lightest nuclei which enable one to study the non-singlet combination of nuclear structure functions 
is the pair of mirror nuclei, $^3$He and $^3$H \cite{Bissey}.

The analysis of deep inelastic scattering (DIS) on nuclear targets demonstrates that the nuclear environment modifies the properties of the nucleons in a number of ways. At small values of Bjorken $x$, the main effects are nuclear shadowing and antishadowing. In this work, we estimate the nuclear shadowing correction to the structure functions, $F_{2}^{^3{\rm He}}(x,Q^2)$ for  $^3$He and  $F_{2}^{^3{\rm H}}(x,Q^2)$ for $^3$H, and for the difference,   $F_{2}^{^3{\rm He}}(x,Q^2)-F_{2}^{^3{\rm H}}(x,Q^2)$, in the region\footnote{
Note that since the transition region between nuclear shadowing and antishadowing  is not constrained well by either models or experiments, we use two models of nuclear shadowing with different cross-over points between the shadowing and antishadowing regions. This fact is reflected in the uncertainty of the upper limit for the shadowing region, $x=0.02 \div 0.045$.} $10^{-4} \le x \le 0.02 \div 0.045$.   
The detailed discussion of our approach to the calculation of nuclear shadowing is presented in  Sect.\ \ref{sec:shadow}.
 For larger values of Bjorken $x$, $0.02 \div 0.045 \le x \le 0.2$, nuclear antishadowing starts to become important. In Sect.\ \ref{sec:anti},
we model antishadowing by requiring the conservation of the number of  valence up and down quarks in $^3$He and $^3$H, which is a generalization of the baryon number sum rule constraint \cite{FS88}.

Our results for small $x$, $x \le 0.2$, can be combined with those of Ref.\ \cite{Bissey} for the large $x$ region in order to present the non-singlet combination $(F_{2}^{^3{\rm He}}(x,Q^2)-F_{2}^{^3{\rm H}}(x,Q^2))/x$ over the full range of Bjorken $x$. 
Sect.\ \ref{sec:results} summarizes our results for two models of shadowing  and two pairs of $^3$He and $^3$H nuclear wave functions.
We also make 
 predictions for the Gottfried integral for the $A$=3 system, defined as \cite{Bissey}
\begin{equation}
I_{G}^{^3{\rm He}, ^3{\rm H}} (z) =\int^{3}_{z} \frac{dx}{x}\Big(F_{2}^{^3{\rm He}}(x,Q^2)-F_{2}^{^3{\rm H}}(x,Q^2)\Big) \ .
\label{def1}
\end{equation}  
In the future, our predictions can be confronted with experiment, for example, with those planned at 
TJLAB \cite{tjlab}, RIKEN \cite{RIKEN1} and RHIC (eRHIC) \cite{Venugopalan}.

\section{Nuclear shadowing correction}
\label{sec:shadow}

The importance of nuclear shadowing in DIS on nuclear targets at small values of Bjorken $x$ is experimentally well-established. For recent reviews of the current situation in experiment and theory, we refer the reader to Refs. \cite{PW}.    
In our approach to nuclear shadowing, we choose to work in the target rest frame, where the dynamics of lepton-nucleus interactions at small $x$ is transparent. 
At small Bjorken $x$, the strong interaction of the virtual photon, emitted by the incident lepton, with hadronic (nucleon or nucleus) targets  takes place in two stages. Firstly, the
photon fluctuates into hadronic configurations $|h_{k} \rangle$ at the 
 distance $l_{c} \approx 1/(2m_{N}x)$ before the target
\begin{equation}
| \gamma^{\ast} \rangle =\sum_{k} |\langle h_{k}|\gamma^{\ast}\rangle|^2 |h_{k} \rangle \ ,
\label{fluct}
\end{equation}
where $|\langle h_{k}|\gamma^{\ast}\rangle|^2$ is the probability that the photon fluctuates into the state $|h_{k} \rangle$.
 In pQCD, the configurations $|h_{k} \rangle$ consist of superpositions of $q \bar{q}$, $q \bar{q} g$, 
$\dots$, Fock states of the virtual photon.  
Secondly, the fluctuations $|h_{k} \rangle$  interact strongly with the target, with some typical hadronic cross sections, $\sigma_{h_{k}A}$. (Here we have chosen the target to be a nucleus with the atomic mass number $A$.)
 Within such a picture, the total virtual  photon-nucleus cross section $\sigma_{\gamma^{\ast}A}$ can be written:
\begin{equation}
\sigma_{\gamma^{\ast}A}=\sum_{k}|\langle h_{k}|\gamma^{\ast}\rangle|^2 \sigma_{h_{k}A} \ .
\label{sh1}
\end{equation}
Here we have suppressed the $x$ and $Q^2$ dependence of $\sigma_{\gamma^{\ast}A}$ for simplicity.
The  $|h_{k} \rangle$-nucleus cross section, $\sigma_{h_{k}A}$, is usually calculated using
 the high-energy scattering formalism of Gribov \cite{Gribov}, which is a generalization
 to high energies  of the Glauber multiple scattering formalism \cite{Glauber}.

The key element of our approach is  an assumption that the sum over the quark-gluon fluctuations of the virtual photon in Eq.\ (\ref{sh1}) can be substituted by some effective state $|h_{eff} \rangle$, which interacts with bound nucleons of the nuclear target with some effective  cross section $\sigma_{eff}$. Examples of the calculation of nuclear shadowing within such an approximation are presented in Refs. \cite{FS88,FGS96,GS99,EPW}.

In the present work, we will use two models for $\sigma_{eff}$. The first model is that of Frankfurt and Strikman \cite{FS99}. The authors 
used the connection between nuclear shadowing in inclusive DIS on nuclei and DIS diffraction in the reaction $\gamma^{\ast}+p \rightarrow X +p^{\prime}$ in order to derive a leading-twist model for  $\sigma_{eff}$. Assuming that higher twist contributions to inclusive DIS are negligible at $Q^2$=4 GeV$^2$, the model of Ref.\ \cite{FS99} gives a 
model-independent\footnote{
The main  assumption which may give a slight model dependence is that  
the strength of multiple rescattering on three or more nucleons is estimated using  the quasi-eikonal approximation. 
Another  assumption, that the nucleus can be describes as a many-nucleon system, is well justified by the small nuclear binding energy per nucleon and also was checked in numerous hadron-nucleus scattering experiments at high energies.}
 description of the main contribution to nuclear shadowing (arising from  virtual photon scattering off two nucleons in the target) in 
nuclear parton densities and structure functions at small Bjorken $x$. For instance, the leading twist contribution to the nuclear shadowing correction  to the deuteron structure function $F_{2}^{d}(x,Q^2)$ can be calculated unambiguously. 
 For nuclei heavier than deuterium, one has to make model-dependent assumptions about $\sigma_{eff}$ for the scattering on three and more nucleons. 
Since the cross-section fluctuations around the average value $\sigma_{eff}$ practically do not affect shadowing \cite{FS99}, one can safely use $\sigma_{eff}$ for the calculation of the virtual photon interaction with more than two bound nucleons and employ
eikonal or quasi-eikonal approximations. $\sigma_{eff}$ of Ref.\ \cite{FS99} as a function of Bjorken $x$ at $Q^2$=4 GeV$^2$ is presented as a solid line in Fig.\ 1.

The second model for $\sigma_{eff}$, which we consider, is based on the two-phase model of nuclear shadowing for inclusive DIS on nuclei of Ref.\ \cite{MT}. This model contains both leading-twist (Pomeron and triple Pomeron) and sub-leading twist (vector meson) contributions to $\sigma_{eff}$. Fig.\ 1 represents the corresponding 
$\sigma_{eff}$ as a function of Bjorken $x$ at $Q^2$=4 GeV$^2$ as a dashed line. 
We note that the difference between $\sigma_{eff}$ of Ref.\ \cite{FS99} (solid line in Fig.\ 1)
and that extracted from Refs. \cite{MT} (dashed line in Fig.\ 1) 
lies both in the inclusion of a
higher twist contribution and  in a different parameterization of the Pomeron contribution.

It is important to note that neither of the  models for  $\sigma_{eff}$ distinguishes between virtual photon rescatterings on protons and neutrons, i.e. $\sigma_{eff}$ is a flavor-singlet cross section. 
In this work, we make a simple extension to the flavor-nonsinglet combination of the virtual photon-nucleon cross sections.

The transition region between nuclear shadowing and antishadowing is poorly known, both experimentally and theoretically. In this region, which approximately lies in the range\footnote{The choice of the lower limit $x=0.02$ is motivated by the model of Ref.\ \cite{FS99}. The upper limit, $x=0.07$, corresponds to the largest Bjorken $x$ for which $F_2^{Ca}/F_2^{D} < 1$ \cite{NMC}.} $0.02 \le x \le 0.07$, nuclear structure functions are modified by a host of nuclear effects. Among these are nuclear shadowing and antishadowing, two-body nucleon-nucleon correlations in the nuclear wave function, the presence of pion degrees of freedom and meson-exchange currents. Since our main emphasis is on the very small Bjorken $x$ region, the detailed description of the nuclear shadowing-antishadowing transition is unimportant.

 Bearing in mind all these nuclear effects, which if ignored, lead to theoretical uncertainties in nuclear structure functions,
we have included in our analysis the shadowing and antishadowing effects only. In addition, 
we have assumed that the calculations of nuclear shadowing, using both models for  $\sigma_{eff}$,  can be performed most reliably  in the range of $10^{-4} \le x \le 0.02$. This explains why the upper limit of $x$ in Fig.\ 1 is set to $x$=0.02. 
Since $\sigma_{eff}$ of Ref.\ \cite{FS99} vanishes at $x$=0.02, we model antishadowing (see Sect.\ \ref{sec:anti}) in the region $0.02 \le x \le 0.2$. On the other hand, the two-phase model \cite{MT} gives $\sigma_{eff}$ which is still quite significant at $x$=0.02 (see Fig.\ 1). In this case, we force  $\sigma_{eff}$ to vanish at $x$=0.045 and make a linear interpolation  between $x$=0.02 and 0.045. In this case, antishadowing is modelled in the region  $0.045 \le x \le 0.2$.

The use of the Gribov-Glauber multiple scattering formalism to calculate $\sigma_{h_{eff}A}$ requires the $|h_{eff} \rangle$-nucleon scattering amplitude and the nuclear wave function.
At high energies, the $|h_{eff} \rangle$-nucleon scattering amplitude $f_{hN}(q)$ is purely imaginary with  good accuracy. Using the optical theorem, $f_{hp}(q)$ for the proton and  $f_{hn}(q)$ for the neutron are related to the total cross sections $\sigma_{eff}^{p}$ and $\sigma_{eff}^{n}$  as 
\begin{eqnarray}
&&f_{hp}(q)=i \sigma_{eff}^{p} e^{-(\beta/2) q^2} \ , \nonumber\\
&&f_{hn}(q)=i \sigma_{eff}^{n} e^{-(\beta/2) q^2} \ 
\label{sh2}
\end{eqnarray}
where $\beta$=6 GeV$^{-2}$ \cite{PW}. Here we have assumed that, in general, the effective cross sections for the interaction with the proton and neutron are different. 

The ground-state wave functions of $^3$He and $^3$H are taken to have  a simple Gaussian form \cite{FGS96,GS99,LS76}
\begin{eqnarray}
&&|\Psi_{^3{\rm He}}|^2 \propto \prod_{l=1}^{l=3} \exp(-\vec{r_{l}}^2/(2 \alpha))\delta^{3}(\sum \vec{r_{l}}) \ , \nonumber\\
&&|\Psi_{^3{\rm H}}|^2 \propto \prod_{l=1}^{l=3} \exp(-\vec{r_{l}}^2/(2 \alpha^{\prime}))\delta^{3}(\sum \vec{r_{l}}) \ .
\label{sh3}
\end{eqnarray}
We have checked that the inclusion of the two-body nucleon-nucleon correlations in the nuclear  wave functions (\ref{sh3}), using the prescription given in Ref.\ \cite{MT}, does not change appreciably the numerical results for nuclear shadowing in the range $10^{-4} \le x \le 0.05$. Hence, we shall employ the wave functions of Eq.\ (\ref{sh3}) in this work.

The nuclear wave functions of Eq.\ (\ref{sh3}) describe the motion of the centers of the nucleons. Thus, the slope parameters $\alpha$ and $\alpha^{\prime}$ should be chosen to reproduce the nuclear matter radii of $^3$He and $^3$H. Assuming that only the proton contributes to the nuclear charge radius, the nuclear matter radius for a nucleus, $R_{m}$, takes the form \cite{BJ77}
\begin{equation}
R_{m}=\sqrt{R_{ch}^2-R_{p}^2} \ ,
\label{rm}
\end{equation}
where $R_{ch}$ and $R_{p}$ are the  charge radii of the nucleus and the proton, $R_{p}=0.880 \pm 0.015$ fm \cite{Rosenfelder}.
In order to estimate the theoretical uncertainty  associated with the nuclear wave functions, we use two values of the average charge radius of $^3$He, 1.976 fm and 1.877 fm, along with the most recent value of the average charge radius of $^3$H, 1.76 fm \cite{AD}. From Eq.\ (\ref{rm}), we obtain the following two pairs of  matter radii of $^3$He and $^3$He: ($R_{m}^{^3{\rm He}}, R_{m}^{^3{\rm H}}$)=(1.769, 1.524) and (1.658, 1.524) fm.
 Using the Gaussian-shaped wave functions (\ref{sh3}) in the   
standard definition of the average nuclear matter radius, one readily finds  that $\alpha=R_{m}^2/2$. This leads to the following two pairs of values for the slopes of the nuclear wave functions of $^3$He and $^3$H (see Eq.\ (\ref{sh3})): 
 ($\alpha$, $\alpha^{\prime}$)=(40.59, 30.06) and (36.11, 30.06) GeV$^{-2}$.
It is important to stress that the fact that $\alpha \neq \alpha^{\prime}$ is a consequence of the charge symmetry breaking in the $^3$He-$^3$H system,
 which is predominantly the Coulomb repulsion in the $^3$He system.
 As will be demonstrated later, this leads to the divergence of the corresponding Gottfried integral.

Using the Gribov-Glauber multiple scattering formalism, along with the elementary scattering amplitude (\ref{sh2}) and the  tri-nucleon ground-state wave functions (\ref{sh3}), one obtains the following $|h_{eff} \rangle$-nucleus ($^3$He and $^3$H) total scattering cross sections
\begin{eqnarray}
&&\sigma_{^3 \rm{He}}=2\sigma_{eff}^{p}+\sigma_{eff}^{n}-\frac{(\sigma_{eff}^{p})^2+2\sigma_{eff}^{p}\sigma_{eff}^{n}}{8 \pi(\alpha+\beta)}e^{-\alpha q_{\parallel}^2}+\frac{(\sigma_{eff}^{p})^2\sigma_{eff}^n}{144 \pi^2(\alpha+\beta)^2} \ , \nonumber\\
&&\sigma_{^3 \rm{H}}=\sigma_{eff}^{p}+2\sigma_{eff}^{n}-\frac{(\sigma_{eff}^{n})^2+2\sigma_{eff}^{p}\sigma_{eff}^{n}}{8 \pi(\alpha^{\prime}+\beta)}e^{-\alpha^{\prime} q_{\parallel}^2}+\frac{(\sigma_{eff}^{n})^2 \sigma_{eff}^p}{144 \pi^2(\alpha^{\prime}+\beta)^2} \ .
\label{c1}
\end{eqnarray}
Here $q_{\parallel}=2m_{N}x$ is the non-zero longitudinal momentum transferred to the target (with $m_{N}$  the nucleon mass). The negligible $x$ dependence of the triple scattering terms (the last terms in the first and second lines of Eqs.\ (\ref{c1})) is omitted.

It is convenient to introduce the flavor singlet, $\sigma_{eff}=(\sigma_{eff}^{p}+\sigma_{eff}^{n})/2$, and flavor non-singlet, $\bar{\sigma}=\sigma_{eff}^{p}-\sigma_{eff}^{n}$, cross sections. 
Note that the two models of $\sigma_{eff}^{p}$ and $\sigma_{eff}^{n}$, those of Refs.\ \cite{FS99} and \cite{MT}, give only the flavor singlet combination $(\sigma_{eff}^{p}+\sigma_{eff}^{n})/2$. Our  analysis will demonstrate that  the leading contribution of the nuclear shadowing correction to the difference $F_{2}^{^3{\rm He}}-F_{2}^{^3{\rm H}}$ is determined by this flavor singlet $\sigma_{eff}$. 
 In this new notation, 
Eqs.\ (\ref{c1}) can be presented as
\begin{eqnarray}
&&\sigma_{^3 \rm{He}}=3\sigma_{eff}+\frac{1}{2}\bar{\sigma}-\frac{3\sigma_{eff}^2+\sigma_{eff}\bar{\sigma}-0.25\bar{\sigma}^2}{8 \pi(\alpha+\beta)}e^{-\alpha q_{\parallel}^2}+\frac{\sigma_{eff}^3+0.5\sigma_{eff}^2\bar{\sigma}-0.25\sigma_{eff}\bar{\sigma}^2-\bar{\sigma}^3/8}{144 \pi^2(\alpha+\beta)^2} \ , \nonumber\\
&&\sigma_{^3 \rm{H}}=3\sigma_{eff}-\frac{1}{2}\bar{\sigma}-\frac{3\sigma_{eff}^2-\sigma_{eff}\bar{\sigma}-0.25\bar{\sigma}^2}{8 \pi(\alpha^{\prime}+\beta)}e^{-\alpha^{\prime} q_{\parallel}^2}+\frac{\sigma_{eff}^3-0.5\sigma_{eff}^2\bar{\sigma}-0.25\sigma_{eff}\bar{\sigma}^2+\bar{\sigma}^3/8}{144 \pi^2(\alpha^{\prime}+\beta)^2} \ .
\label{c2}
\end{eqnarray}

\begin{figure}
\epsfig{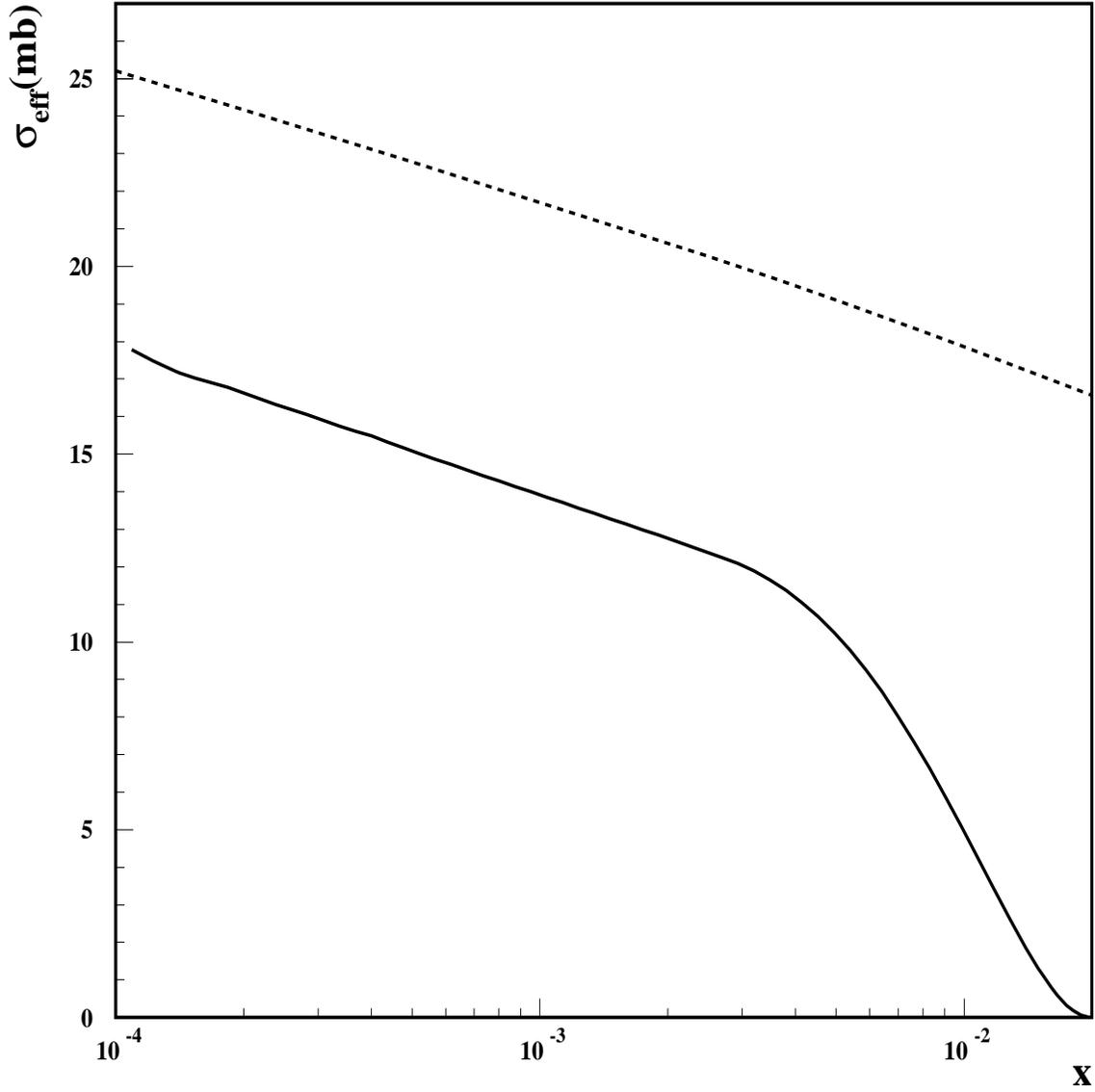}
\vspace{1cm}
\caption{$\sigma_{eff}$ as a function of Bjorken $x$ at $Q^2$=4 GeV$^2$ from Refs.\ [23] (solid curve) and [24] (dashed curve).
Note that the dashed curve includes a higher twist contribution and an earlier parameterization of the Pomeron.
}
\label{fig:sigma}
\end{figure}

It is useful to introduce the short-hand notation
\begin{eqnarray}
&&f_{\alpha}=\frac{\sigma_{eff}}{8 \pi(\alpha+\beta)}e^{-\alpha q_{\parallel}^2} \ , \nonumber\\
&&g_{\alpha}=\frac{\sigma_{eff}^2}{144 \pi^2(\alpha+\beta)^2} \ , \nonumber\\
&&f_{\alpha^{\prime}}=\frac{\sigma_{eff}}{8 \pi(\alpha^{\prime}+\beta)}e^{-\alpha^{\prime} q_{\parallel}^2} \ , \nonumber\\
&&g_{\alpha^{\prime}}=\frac{\sigma_{eff}^2}{144 \pi^2(\alpha^{\prime}+\beta)^2} \ .
\label{c3}
\end{eqnarray}
Note that $f_{\alpha}$, $g_{\alpha}$, $f_{\alpha^{\prime}}$ and $g_{\alpha^{\prime}}$ are functions of $x$. Their $x$ dependence originates predominantly from the $x$ dependence of $\sigma_{eff}$, see Fig.\ \ref{fig:sigma}. 
There is an additional $x$ dependence
from  the non-zero value of $q_{\parallel}$, which becomes important for $x \gtrsim 0.05$.

Using the short-hand notation of Eqs.\ (\ref{c3}) and ignoring 
the
terms of the order of $\bar{\sigma}^2$ and $\bar{\sigma}^3$, Eqs.\ (\ref{c2}) become
\begin{eqnarray}
&&\sigma_{^3 \rm{He}}=3\sigma_{eff}(1-f_{\alpha}+g_{\alpha}/3)+\frac{\bar{\sigma}}{2}(1-2f_{\alpha}+g_{\alpha}) \ , \nonumber\\
&&\sigma_{^3 \rm{H}}=3\sigma_{eff}(1-f_{\alpha^{\prime}}+g_{\alpha^{\prime}}/3)-\frac{\bar{\sigma}}{2}(1-2f_{\alpha^{\prime}}+g_{\alpha^{\prime}}) \ .
\label{c4}
\end{eqnarray}
It is important to stress that Eqs.\ (\ref{c4}) demonstrate that nuclear shadowing in the non-vacuum channel (the coefficient in front of the $f_{\alpha}\bar{\sigma}$ ($f_{\alpha^{\prime}}\bar{\sigma}$) term) is twice as large as that in the vacuum channel (the coefficient in front of the $f_{\alpha} \sigma_{eff}$ ($f_{\alpha^{\prime}} \sigma_{eff}$) term). This was first suggested in Ref. \cite{FS88}. 
A similar conclusion was reached in the analysis of polarized DIS on $^3$He \cite{FGS96} and $^7$Li \cite{GS99}.
The observation that nuclear shadowing is enhanced by a factor of 2 in the non-vacuum channel, as compared to the vacuum channel, 
seems to be a generic property of nuclear shadowing and it requires more theoretical work.

Introducing the  structure functions $F_2(x,Q^2)$ as
\begin{eqnarray}
&&F_{2}^{^3{\rm He}}(x,Q^2) \propto \sigma_{^3 \rm{He}} \ , \nonumber\\
&&F_2^p(x,Q^2)+F_2^n(x,Q^2) \propto \sigma_p+\sigma_n = 2 \sigma_{eff} \ , \nonumber\\
&&F_2^p(x,Q^2)-F_2^n(x,Q^2) \propto \sigma_p-\sigma_n = \bar{\sigma} \ , 
\end{eqnarray}
one can write for the structure functions of $^3$He and $^3$H in the shadowing region of Bjorken $x$ as
\begin{eqnarray}
&&F_{2}^{^3{\rm He}}=2F_{2}^{p}+F_{2}^{n}-F_{2}^{p}(2.5f_{\alpha}-g_{\alpha})-F_{2}^{n}(0.5f_{\alpha}) \ , \nonumber\\
&&F_{2}^{^3{\rm H}}=2F_{2}^{n}+F_{2}^{p}-F_{2}^{n}(2.5f_{\alpha^{\prime}}-g_{\alpha^{\prime}})-F_{2}^{p}(0.5f_{\alpha^{\prime}}) \ .
\label{sh4}
\end{eqnarray}
In Eq.\ (\ref{sh4}), the obvious $x$ and $Q^2$ dependence of the structure functions has been suppressed.  
Eqs.\ (\ref{sh4}) describe the modification of  $F_{2}^{^3{\rm He}}(x,Q^2)$ and $F_{2}^{^3{\rm H}}(x,Q^2)$  at small Bjorken $x$, as a consequence of nuclear shadowing.
 We observe a qualitatively new effect -- the violation of SU(2) isospin (charge) symmetry in the wave functions of the $A=3$ system, which enters through the shadowing correction, induces a violation of  SU(2) isospin symmetry for the structure functions $F_{2}^{^3{\rm He}}$ and $F_{2}^{^3{\rm H}}$.
The latter means that $F_{2}^{^3{\rm He}}$ and $F_{2}^{^3{\rm H}}$ are no longer related by a rotation in the isospin space. In other words, the charge symmetry violation in the wave functions of the $A=3$ system results in the SU(2) isospin symmetry breaking for nuclear shadowing (regardless of the fact that nuclear shadowing is determined by the SU(2)-symmetric exchange with vacuum quantum numbers (Pomeron)).
 
As explained above, we assume that using Eq.\ (\ref{sh4}), nuclear shadowing can be calculated most reliably in the range $10^{-4} \le x \le 0.02$. At higher Bjorken $x$, nuclear antishadowing begins to play a role. Our model-dependent treatment of the antishadowing contribution is presented in next section.

\section{Nuclear antishadowing correction}
\label{sec:anti}

The dynamical mechanism of antishadowing is unknown. Thus, at the present stage, all considerations of nuclear antishadowing are model-dependent. One possible approach to modelling nuclear antishadowing uses the baryon number and momentum sum rules \cite{FS88,FLS,Eskola}. 
The authors of Refs.\ \cite{FS88} suggest the following scenario, which is consistent with  the data. 
(However, it follows from the data only if an assumption is made that 
higher twist effects are small.
This assumption is very natural for the case of  Drell-Yan data and also supported by approximate scaling of DIS data).
Nuclear shadowing is present in the valence quark, sea quark and gluon parton densities; nuclear antishadowing is present only in the valence and gluon parton densities.

Using Eqs.\ (\ref{sh4}) we can calculate the nuclear quark parton densities.  Adding the antishadowing contribution to the valence quarks, this leads to:
\begin{eqnarray}
&&u^{^3{\rm He}}_{val}=2u_{val}+d_{val}+(2.5u_{val}+0.5d_{val})(-f_{\alpha}+f_{\alpha,u}^{anti})+u_{val}g_{\alpha} \ , \nonumber\\
&&d^{^3{\rm He}}_{val}=2d_{val}+u_{val}+(2.5d_{val}+0.5u_{val})(-f_{\alpha}+f_{\alpha,d}^{anti})+d_{val}g_{\alpha} \ , \nonumber\\
&&{\bar u^{^3{\rm He}}}=2\bar{u}+\bar{d}+(2.5\bar{u}+0.5\bar{d})(-f_{\alpha})+\bar{u}g_{\alpha} \ , \nonumber\\
&&{\bar d^{^3{\rm He}}}=2\bar{d}+\bar{u}+(2.5\bar{d}+0.5\bar{u})(-f_{\alpha})+\bar{d}g_{\alpha} \ . 
\label{b1}
\end{eqnarray}
where $u_{val}$ and $d_{val}$ stand for the valence up and down quark parton densities. 
The unknown functions $f_{\alpha,u}^{anti}$ and $f_{\alpha,d}^{anti}$ describe nuclear antishadowing for the valence up and down quarks in $^3$He.
In order to obtain  nuclear quark parton densities in $^3$H, one needs to replace $\alpha$ by $\alpha^{\prime}$ and $u$ by $d$ in 
the right hand side of
Eqs.\ (\ref{b1}).

In order to find the functions $f_{\alpha,u}^{anti}$ and $f_{\alpha,d}^{anti}$, we used conservation of valence up and down quarks in $^3$He
\begin{eqnarray}
&&\int_{0}^{3} dx u^{^3{\rm He}}_{val}(x)=\int_{0}^{3} dx (2u_{val}(x)+d_{val}(x)) \ , \nonumber\\
&&\int_{0}^{3} dx d^{^3{\rm He}}_{val}(x)=\int_{0}^{3} dx (2d_{val}(x)+u_{val}(x)) \ .
\label{sh12}
\end{eqnarray}
The corresponding sum rules are valid for 
$u^{^3{\rm H}}_{val}$ and  $d^{^3{\rm H}}_{val}$ in  $^3$H after the replacement $u \leftrightarrow d$ in 
the right hand side of
 Eqs.\ (\ref{sh12}).

Substituting the first two of Eqs.\ (\ref{b1}) into (\ref{sh12}), one obtains the following constraint on $f_{\alpha,u}^{anti}$ and $f_{\alpha,d}^{anti}$
\begin{eqnarray}
&&\int_{0.0001}^{x_0} dx \Big((2.5u_{val}(x)+0.5d_{val}(x))f_{\alpha}-u_{val}(x)g_{\alpha}\Big)=\int_{x_0}^{0.2} dx \Big(2.5u_{val}(x)+0.5d_{val}(x)\Big)f_{\alpha,u}^{anti} \ , \nonumber\\
&&\int_{0.0001}^{x_0} dx \Big(2.5d_{val}(x)+0.5u_{val}(x))f_{\alpha}-d_{val}(x)g_{\alpha}\Big)=\int_{x_0}^{0.2} dx \Big(2.5d_{val}(x)+0.5u_{val}(x)\Big)f_{\alpha,d}^{anti} \ ,
\label{sh13}
\end{eqnarray} 
where $x_0$=0.02 for the calculations with $\sigma_{eff}$ of Ref.\ \cite{FS99} and $x_0$=0.045 
for the calculations with $\sigma_{eff}$ based on Refs.\ \cite{MT}. 
In the latter case, we have assumed that $\sigma_{eff}$ linearly decreases from $x$=0.02 and becomes zero at $x_0$=0.045. This choice of  $x_0$ is motivated by the NMC data on $^4$He \cite{NMC} since $F_{2}^{\rm{He}}/F_{2}^{\rm{D}}=1$ at $x$=0.045.

Using Eqs.\ (\ref{b1}) for $^3$He and the corresponding equations for $^3$H, we obtain the following equations for the nuclear structure functions
\begin{eqnarray}
&&F_{2}^{^3{\rm He}}=2F_{2}^{p}+F_{2}^{n}-F_{2}^{p}(2.5f_{\alpha}-g_{\alpha})-F_{2}^{n}(0.5f_{\alpha}) \nonumber\\
&&+\frac{1}{9}\Big(F_{2val}^{p}(\frac{114}{5}f_{\alpha,u}^{anti}-\frac{3}{10}f_{\alpha,d}^{anti})+F_{2val}^{n}(-\frac{6}{5}f_{\alpha,u}^{anti}+\frac{57}{10}f_{\alpha,d}^{anti})\Big) \ , \nonumber\\
&&F_{2}^{^3{\rm H}}=2F_{2}^{n}+F_{2}^{p}-F_{2}^{n}(2.5f_{\alpha^{\prime}}-g_{\alpha^{\prime}})-F_{2}^{p}(0.5f_{\alpha^{\prime}}) \nonumber\\
&&+\frac{1}{9}\Big(F_{2val}^{n}(\frac{114}{5}f_{\alpha^{\prime},d}^{anti}-\frac{3}{10}f_{\alpha^{\prime},u}^{anti})+F_{2val}^{p}(-\frac{6}{5}f_{\alpha^{\prime},d}^{anti}+\frac{57}{10}f_{\alpha^{\prime},u}^{anti})\Big) \ ,
\label{b2}
\end{eqnarray}
where $F_{2\, val}^{p}$ and $F_{2\, val}^{n}$ are the structure functions including only valence quarks.
Eqs.\ (\ref{b2}) describe the nuclear shadowing and antishadowing corrections to the nuclear structure functions $F_{2}^{^3{\rm He}}$ and $F_{2}^{^3{\rm H}}$
 over the range $10^{-4} \le x \le 0.2$.

We would like to stress again that
as one can see from Eqs.\ (\ref{b1}) and (\ref{b2}), the violation of charge symmetry in the tri-nucleon  wave functions induces  SU(2) isospin symmetry breaking in the quark parton densities and structure functions. In particular, one finds from Eq.\ (\ref{b1}) that $u^{^3{\rm He}} \neq d^{^3{\rm H}}$, and from Eq.\ (\ref{b2}) that $F_{2}^{^3{\rm He}}$ is not related to $F_{2}^{^3{\rm H}}$ by the permutation $p \leftrightarrow n$.

 The novelty of Eqs.\ (\ref{b2}) consists in the fact that they present the nuclear shadowing and antishadowing corrections to structure functions which by themselves are neither flavor singlet nor flavor non-singlet.   
 Until now, all analyses of nuclear shadowing in DIS on nuclei were concerned with nuclei with an equal number of protons and neutrons -- i.e., flavor singlet nuclei. In applying the previously developed theory of nuclear shadowing and antishadowing to DIS on $^3$He and $^3$H and deriving Eqs.\ (\ref{b2}), we have implicitly made the following assumptions 
for the non-singlet combinations of the structure functions $F_{2}$ and quark densities.
  We have assumed that $\sigma_{eff}$, which controls the amount of nuclear shadowing is the same for $u-d$ and $\bar{u}+\bar{d}$. In other words, in Eq.\ (\ref{c4}), 
the same $\sigma_{eff}$ determines the shadowing correction to $\sigma_{eff}$ (first terms) and $\bar{\sigma}$ (second term).
Another assumption was that  antishadowing for $\bar{u}$ and $\bar{d}$ is the same as for $\bar{u}+\bar{d}$ -- i.e.,
it is nil. 
We believe that regardless of the model-dependent nature of our estimates, Eqs.\ (\ref{b2}) provide a reasonable  estimate of  the low $x$ nuclear corrections to the  structure functions  $F_{2}^{^3{\rm He}}$ and $F_{2}^{^3{\rm H}}$.

\section{Results and discussions}
\label{sec:results}

Equations \ (\ref{b2}) have been used to predict the difference $(F_{2}^{^3{\rm He}}-F_{2}^{^3{\rm H}})/x$ as a function of $x$, in the range of $10^{-4} \le x$. In order to test the sensitivity to the input parameters, we have considered 5 following combinations of the slopes of $^3$He and $^3$H ground-state wave functions (\ref{sh3}) and models of nuclear shadowing:

\begin{enumerate}

\item{$\alpha$=40.59 GeV$^{-2}$, $\alpha^{\prime}$=30.06 GeV$^{-2}$, $\sigma_{eff}$ of Ref.\ \cite{FS99} with $x_{0}$=0.02 ($x_0$ is the point of the transition from shadowing to antishadowing, i.e. $f_{\alpha,u}(x_0)=f_{\alpha,d}(x_0)=f_{\alpha,u}^{anti}(x_0)=f_{\alpha,d}^{anti}(x_0)=0$ and $f_{\alpha^{\prime},u}(x_0)=f_{\alpha^{\prime},d}(x_0)=f_{\alpha^{\prime},u}^{anti}(x_0)=f_{\alpha^{\prime},d}^{anti}(x_0)=0$.
The parameter $x_0$ enters through Eqs.\ (\ref{sh13}));}

\item{$\alpha$=36.11 GeV$^{-2}$, $\alpha^{\prime}$=30.06 GeV$^{-2}$, $\sigma_{eff}$ of Ref.\ \cite{FS99} with $x_{0}$=0.02;}

\item{$\alpha$=40.59 GeV$^{-2}$, $\alpha^{\prime}$=30.06 GeV$^{-2}$, $\sigma_{eff}$ of Refs.\ \cite{MT} with $x_{0}$=0.045;}
 
\item{$\alpha$=36.11 GeV$^{-2}$, $\alpha^{\prime}$=30.06 GeV$^{-2}$, $\sigma_{eff}$ of Refs.\ \cite{MT} with $x_{0}$=0.045;}

\item{$\alpha$=40.59 GeV$^{-2}$, $\alpha^{\prime}$=40.59 GeV$^{-2}$, $\sigma_{eff}$ of Ref.\ \cite{FS99} with $x_{0}$=0.02.}

\end{enumerate}

For each of these cases, we have assumed the following simple shapes of $f_{\alpha,u}^{anti}$, $f_{\alpha,d}^{anti}$, $f_{\alpha^{\prime},u}^{anti}$ and  $f_{\alpha^{\prime},d}^{anti}$. (Below we present only $f_{\alpha,u}^{anti}$, with  the others being defined in a similar way.) 
\begin{eqnarray}
&&f_{\alpha,u}^{anti}=\frac{h_u}{0.09-x_0}(x-x_0), \,  x_0 \le x \le 0.09 \ , \nonumber\\
&&f_{\alpha,u}^{anti}=\frac{h_u}{0.11}(0.2-x), \  0.09 \le x \le x_0 \ , \nonumber\\
&&f_{\alpha,u}^{anti}(x)=0, \ {\rm elsewhere} \ .
\label{sh15}
\end{eqnarray} 
The most recent NMC data on $F_{2}^{^4{\rm He}}/F_{2}^{D}$ indicates that the antishadowing contribution peaks at $x=0.09$ \cite{NMC}. Since DIS on $^3$He or $^3$H has not been measured, we assumed that antishadowing also peaks at $x=0.09$ for DIS on $^3$He or $^3$H. This fact is reflected in the parameterization of Eqs.\ (\ref{sh15}).  
The constants $h$ are chosen so that Eqs.\ (\ref{sh13}) are satisfied.

For quark parton densities in the proton we used the leading order CTEQ5 parameterization CTEQ5L \cite{CTEQ}. Note also that throughout our work we use the leading order  expression for the structure functions 
$F_2(x,Q^2)$. 
This allows us 
to omit an explicit consideration of  gluons and forces us to use the leading order quark parton densities, such as for example CTEQ5L.

Figures 2-5 present $(F_{2}^{^3{\rm He}}-F_{2}^{^3{\rm H}})/x$ as a function of Bjorken $x$ at $Q^2$=4 GeV$^2$ for the 5 combinations of $(\alpha,\alpha^{\prime})$ and $\sigma_{eff}$ given above. The solid lines are results of the calculations using Eqs.\ (\ref{b2}) over the range  $10^{-4} \le x \le 0.2$. At larger $x$, the calculations of Saito {\it et al.} \cite{Bissey} were used. For each of the 5 cases, the solid lines should be compared to the dotted lines, which present  $(F_{2}^{^3{\rm He}}-F_{2}^{^3{\rm H}})/x$ in the absence of all nuclear effects, when $(F_{2}^{^3{\rm He}}-F_{2}^{^3{\rm H}})/x=(F_{2}^{p}-F_{2}^{n})/x$.

The dotted line in Fig.\ 2 presents $(F_{2}^{^3{\rm He}}-F_{2}^{^3{\rm H}})/x$ for case 5, when the slopes $\alpha$ and $\alpha^{\prime}$ are chosen to be equal. The large difference between the solid and dotted lines at small $x$ demonstrates 
that the rise of $(F_{2}^{^3{\rm He}}-F_{2}^{^3{\rm H}})/x$ at small $x$ 
originates from non-cancellation of divergent terms in the flavor nonsinglet combination  of structure functions $F_2$ of the bound proton and neutron,
 when $\alpha \neq \alpha^{\prime}$.
This result implies that charge symmetry breaking (in the present case, mainly from the  Coulomb force) is very important and enhances the difference of the structure functions of mirror nuclei at small $x$.

In order to better appreciate the magnitude of these nuclear effects (nuclear shadowing and antishadowing) for the flavor non-singlet combinations of structure functions, it should be compared to the contribution of nuclear shadowing and antishadowing to the singlet combinations of structure functions. Figure 6 presents the flavor singlet combinations  $F_{2}^{^3{\rm He}}+F_{2}^{^3{\rm H}}$ (solid and dashed lines) and  $3(F_{2}^{p}+F_{2}^{n})$ (dotted line) as functions of $x$ at $Q^2$=4 GeV$^2$. For the solid and dashed lines we used the first and third combinations of  $(\alpha,\alpha^{\prime})$ and $\sigma_{eff}$ described in the text above\footnote{Note also that when  $\sigma_{eff}$ is fixed, the variation of $(\alpha,\alpha^{\prime})$ leads to very insignificant changes in the amount of shadowing and antishadowing. Thus, the solid line in Fig.\ 6 corresponds to combinations 1,2 and 5; the dashed line  corresponds to combinations 3 and 4.}. One can see from Fig.\ 6 that, in contrast to the flavor non-singlet structure functions, 
 nuclear shadowing decreases $F_{2}^{^3{\rm He}}+F_{2}^{^3{\rm H}}$ as compared to $3(F_{2}^{p}+F_{2}^{n})$ but this effect is not as dramatic. The decrease is 4.5\% (6\%) for the solid (dashed) line at $x=10^{-4}$.
The main conclusion, which one can draw from comparing Figs.\ 2, 3, 4, 5 to Fig.\ 6, is that
because of the charge symmetry breaking in the nuclear ($^3$He and $^3$H) wave functions, 
 the nuclear shadowing correction is much more significant for the flavor non-singlet combination of structure functions, $(F_{2}^{^3{\rm He}}-F_{2}^{^3{\rm H}})/x$, than for the flavor singlet combination $F_{2}^{^3{\rm He}}+F_{2}^{^3{\rm H}}$. 
For example, for case 3 and at $x=10^{-4}$ and $Q^2$=4 GeV$^2$, 
$(F_{2}^{^3{\rm He}}-F_{2}^{^3{\rm H}})/(F_{2}^{^3{\rm He}}+F_{2}^{^3{\rm H}})$=0.0085. However, it should be noticed that since the effect of nuclear shadowing in parton densities decreases
because of the QCD evolution, 
we  expect the ratio $(F_{2}^{^3{\rm He}}-F_{2}^{^3{\rm H}})/(F_{2}^{^3{\rm He}}+F_{2}^{^3{\rm H}})$
 to decrease as $Q^2$ increases.

We used our results for $(F_{2}^{^3{\rm He}}-F_{2}^{^3{\rm H}})/x$ in order to
investigated the role played by the small $x$ nuclear effects on the Gottfried integral. Table 1 present our estimates of the Gottfried integral $I_{G}^{^3{\rm He}, ^3{\rm H}} (10^{-4})$, defined by Eq.\ (\ref{def1}), and the ratio $I_{G}^{^3{\rm He}, ^3{\rm H}} (10^{-4})/I_{G}^{p,n} (10^{-4})$ ($I_{G}^{p,n} (10^{-4})$ is the Gottfried integral for the free proton and neutron. We obtained $I_{G}^{p,n} (10^{-4})=0.24$ using CTEQ5L, which is in a good agreement with the NMC result $I_{G}^{p,n} (10^{-4})=0.235 \pm 0.026$ \cite{NMC}.)
We found that the effect of nuclear shadowing increases the  Gottfried integral for the $^3$He-$^3$H system by $11 \div 36$ \%, depending on the combination 
($\alpha$, $\alpha^{\prime}$) and $\sigma_{eff}$.

So far we discussed small, but finite, Bjorken $x$ behaviour of $(F_{2}^{^3{\rm He}}-F_{2}^{^3{\rm H}})/x$ and the integral  thereof. At least from the theoretical  point of view, one can ask the question: what happens to $I_{G}^{^3{\rm He}, ^3{\rm H}} (x)$ when $x \rightarrow 0$? Our analysis seems to suggest that the Gottfried integral for the $^3$He-$^3$H system is divergent logarithmically because of the non-cancellation of the factor $1/x$. We observe that this result is not paradoxical since the Gottfried integral is not constrained by current algebra -- as, for example, the Bjorken sum rule. Thus, the value of the Gottfried integral is not related to any physical observable or constant and, in principle, can be infinite.

Our statement that $I_{G}^{^3{\rm He}, ^3{\rm H}} (0)$ diverges is supported
 by the analysis of the total virtual photon-nucleus cross section (structure function $F_{2}$)
 at small values of Bjorken $x$  within the Gribov model \cite{Gribov}. 
Indeed, for DIS on nucleon or nucleus,  one can write the dispersion integral\footnote{In general, one has to use the double dispersion representation. However, in the black body limit discussed here (see Eq.\ (\ref{disp2})), only diagonal transitions contribute \cite{Gribov}.}
  over diffractive masses $M$  for the structure function $F_{2}$:
\begin{equation}
F_{2}=\frac{Q^2}{12 \pi^3} \int^{M_{max}^2}_{0} \frac{dM^2 \rho(M^2) M^2 \sigma(M^2)}{(M^2+Q^2)^2} \ . 
\label{disp1}
\end{equation}
Here $\rho(M^2)$ is the the ratio $\sigma(e^{+}e^{-} \rightarrow hadrons)/\sigma(e^{+}e^{-} \rightarrow \mu^{+} \mu^{-})$ ($M^2$ being mass squared of the produced final hadronic state, denoted by ``{\it hadrons}''); $\sigma(M^2)$ is the photon-target cross section for the production of the final state with mass squared $M^2$.
The key assumption of the  model is that
 when $Q^2$ is constant, $x$ is very small and $A \rightarrow \infty$ (very heavy nuclear target), the hadronic fluctuations of the virtual photon $|h_{k} \rangle$ (see Eq.\ (\ref{fluct})) interact with the nucleus with the maximal possible cross section $2 \pi R_{A}^2$ ($R_{A}$ is the size of the nucleus) \cite{Gribov}. This set of approximations is sometimes called the black body limit.
Thus, the contribution of the range of $M^2$ for which the black body limit holds, i.e. $\sigma(M^2)=2 \pi R_{A}^2$, to the structure function $F_{2}$ is 
\begin{equation}
F_{2}=\frac{Q^2 2 \pi R_{A}^2}{12 \pi^3} \int^{M_{max}^2}_{0} \frac{dM^2 \rho(M^2) M^2}{(M^2+Q^2)^2} \ . 
\label{disp2}
\end{equation}
The upper limit of integration, $M_{max}^2$, is defined as the maximal mass squared of a diffractively produced intermediate state when, at fixed $x$ and $Q^2$, the black body limit is reached for all essential fluctuations of the virtual photon. Within the dipole picture of pQCD, it was estimated in Ref.\ \cite{MFGS} that 
$M_{max}^2=Q^2 x_{bbl}/x$, where $x_{bbl}$ is the critical Bjorken $x^{\prime}$ entering the dipole formulation of Ref.\ \cite{MFGS}, when the black body limit is achieved.  The factor $x_{bbl}$ depends on the details of a particular dipole model and, in general, significantly affects the absolute value of $F_{2}$ predicted by Eq.\ (\ref{disp2}). Since we are concerned with qualitative and model-independent aspects of the $x$ behaviour of $F_2$ following from Eq.\ (\ref{disp2}), after taking the integral over the diffractive masses in Eq.\ (\ref{disp2}), we can present the nuclear structure function  in the form
\begin{equation}   
F_{2} \propto Q^2 R_{A}^2 \ln(1/x) \ + {\it subleading \ terms} \ .
\label{disp3}
\end{equation}

 The application of the black body limit as $x \rightarrow 0$ 
is also justified for light nuclei and nucleons. In particular, using Eq.\ (\ref{disp3}), one obtains for the difference of the structure functions
of $^3$He and $^3$H:
\begin{equation}
F_{2}^{^3{\rm He}}(x,Q^2)-F_{2}^{^3{\rm H}}(x,Q^2)=Q^2 (R^2_{^3{\rm He}}-R^2_{^3{\rm H}})\ln(1/x) \ + {\it subleading \ terms}
.
\label{u2}
\end{equation} 
The charge symmetry breaking in the $^3$He-$^3$H system manifests itself as the non-equality of charge and, hence, nuclear matter radii of $^3$He ($R_{^3{\rm He}}$)  and $^3$H ($R_{^3{\rm H}}$). 
Substituting Eq.\ (\ref{u2}) into the Gottfried integral yields an integral, divergent as $(\ln(1/x))^2$ as $x \rightarrow 0$. Hence, we conclude that our analysis of nuclear shadowing and the one within the framework of the black body approximation shows that the Gottfried integral for the $^3$He-$^3$H system is divergent.

It is interesting to note that the above discussed phenomena should also be relevant for  the free proton and neutron. In this case, similarly to the tri-nucleon system, small charge symmetry breaking makes the (hadronic) sizes of the proton and neutron different. 
Specifically, two effects work in the direction of making the radius of the proton larger than the radius of the neutron. These are the Coulomb repulsion and the quark mass difference. (Since the neutron consists of two $d$ quarks and one $u$ quark and the $d$ quark is heavier than the $u$, the size of the neutron is smaller than that of  the proton consisting of two $u$ quarks and one $d$ quark.)
The difference in sizes should lead to different photoabsorption cross sections on the proton and neutron.

The analysis of  (virtual) photon-hadron interactions demonstrated that in order to successfully describe  the data, the photon should 
 contain ``soft'' and ``hard'' contributions. The soft part interacts with the target with some typical hadronic cross section. Phenomenologically, cross sections of soft interactions are proportional to the square of the radius of the target hadron -- see e.g. Ref.\ \cite{HP}. In light of the argument presented above for the size of the valence quark distributions in the proton and neutron,
the soft component of the photon should interact with the proton with a larger cross section.
One can expect  a similar effect for the hard component of the photon. The hard cross section is proportional to the gluon field of the target with a cutoff proportional to the size of the target. This makes the cross section for interaction with the proton larger  than that with the neutron. Hence, in the limit of very small values of Bjorken $x$, the total photoabsorption cross section on the proton is larger than on the neutron. In other words, we expect that $F_{2}^{p}$ should be greater than  $F_{2}^{n}$, which          
would  lead to the divergence of the Gottfried integral,  $I_{G}^{p,n} (0)$.
 Further investigations of this interesting question are necessary. If, indeed, $I_{G}^{p,n} (0)$ is infinite, modern parton 
 distributions  need to be revised since they impose the condition $F_{2}^{p}-
F_{2}^{n} \to 0$ as $x \to 0$ and, hence, 
give a finite value of $I_{G}^{p,n} (0)$.

\section{Conclusion}

We considered the influence of the nuclear effects of shadowing and antishadowing on the structure functions  $F_{2}^{^3{\rm He}}$ of $^3$He and $F_{2}^{^3{\rm H}}$ of $^3$H. We found that these nuclear effects increase the Gottfried integral $I_{G}^{^3{\rm He}, ^3{\rm H}} (10^{-4})$ by $11 \div 36$\%, depending on the model used for the nuclear wave functions and for the calculation of nuclear shadowing.
 We observed that the violation of charge symmetry for the nuclear wave functions of $^3$He and $^3$H induces charge symmetry breaking for the nuclear quark parton densities (as a result of  the nuclear shadowing correction).  
This leads to the conclusion that 
 the Gottfried integral, integrated over the whole region of Bjorken $x$, $I_{G}^{^3{\rm He}, ^3{\rm H}} (0)$, is divergent. 
It is expected that even in the case of the free nucleon, 
the hadronic sizes of the proton and neutron should be different because of the small 
charge symmetry breaking effect.  This suggests that the 
Gottfried integral of the free nucleon should be  divergent at very small $x$. 
It will be very interesting to study the Gottfried integral of the free nucleon 
at very small $x$ in the future.

Experiments on
 DIS off mirror nuclei with large isospin asymmetry should be 
possible in the future \cite{tjlab,RIKEN1,Venugopalan}.  The observation of some 
deviation from the present calculations 
would provide 
information on phenomena involving non-pQCD
dynamics
 (like 
the pion fields) in a nuclear medium.  
If one
 could vary the atomic number ($A$) and the difference 
between the proton and neutron numbers ($Y=Z-N$) independently in measuring  
the nuclear structure functions of unstable mirror nuclei \cite{Bissey}, 
it would stimulate a great deal of work which may eventually lead to 
genuinely new information on the dynamics of nuclear systems.  

\section{Acknowledgements}
%\vspace{1cm}
%\noindent Acknowledgement: \\

We would like to thank F. Bissey for allowing us to use his result of the 
structure functions of $^3$He and $^3$H in the large $x$ region. 
One of us (M.S.) would also like to thank L. Frankfurt for useful discussions of the Gottfried integral for the free nucleon case. 
 This work 
was supported by the Australian Research Council and U.S. Department of Energy.

\begin{table}
\begin{tabular}{|c|c|c||}
\hline
Case number & $I_{G}^{^3{\rm He}, ^3{\rm H}} (10^{-4})$ & $I_{G}^{^3{\rm He}, ^3{\rm H}} (10^{-4})/I_{G}^{p,n} (10^{-4})$ \\
\hline 
 1 & 0.286 & 1.19 \\
 2 & 0.267 & 1.11 \\
 3 & 0.328 & 1.36 \\
 4 & 0.296 & 1.23 \\
 5 & 0.242 & 1.01 \\
\hline
\end{tabular}

\vspace{1 cm}

\caption{The Gottfried integral $I_{G}^{^3{\rm He}, ^3{\rm H}} (10^{-4})$, defined by Eq.\ (\ref{def1}), and the ratio of nuclear and free space Gottfried integrals $I_{G}^{^3{\rm He}, ^3{\rm H}}(10^{-4})/I_{G}^{p,n} (10^{-4})$ for the 5 combinations of ($\alpha, \alpha^{\prime}$) and $\sigma_{eff}$ described in the text.} 

\label{table:one}
\end{table}

\begin{figure}
\epsfig{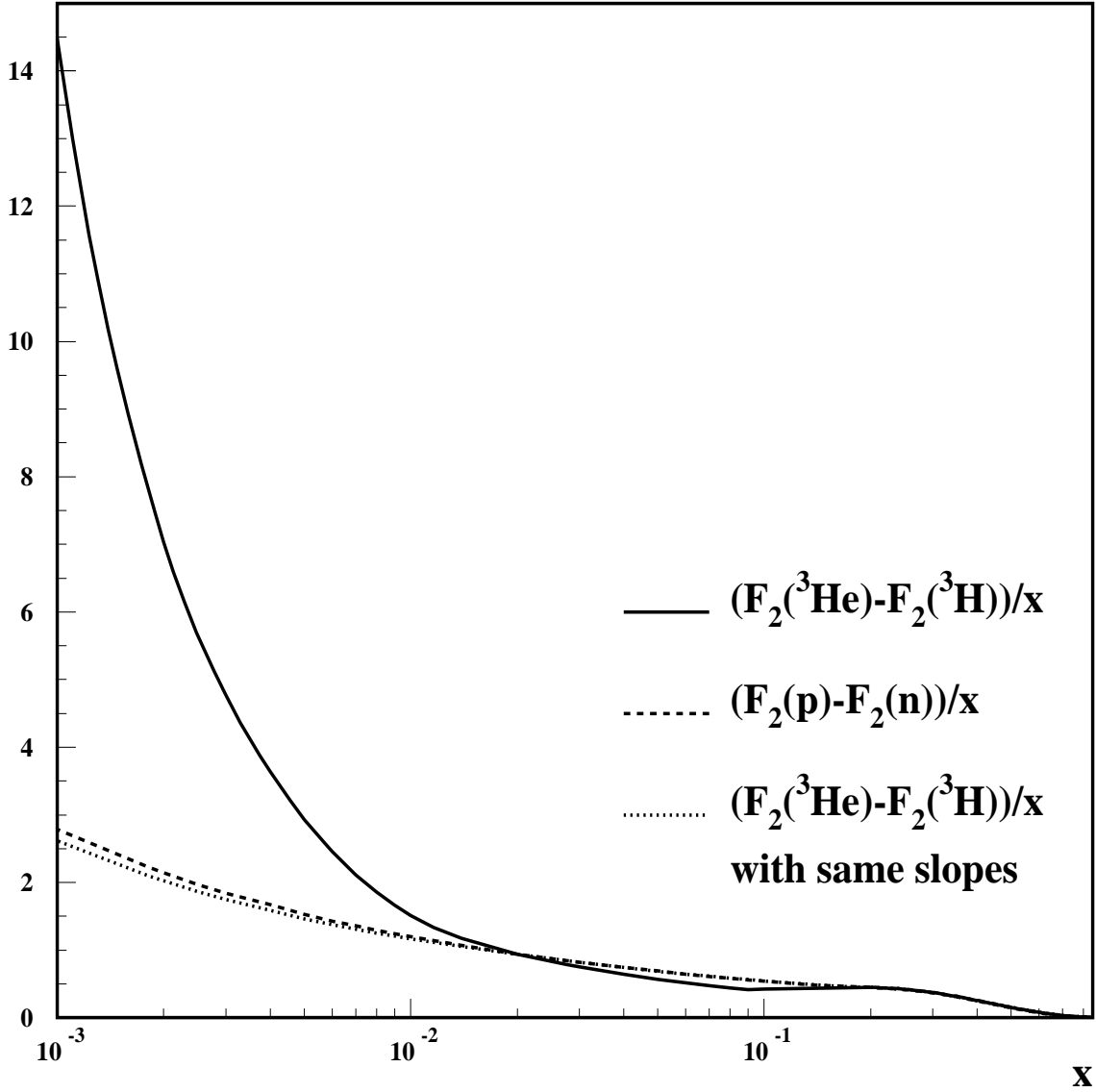}
\label{fig:v1}
\vspace{1cm}
\caption{Cases 1 and 5. The solid (dashed) line  represents $(F_{2}^{^3{\rm He}}-F_{2}^{^3{\rm H}})/x$ ($(F_{2}^{p}-F_{2}^{n})/x$) as a function of Bjorken $x$ at $Q^2$=4 GeV$^2$.
Case 5 for $(F_{2}^{^3{\rm He}}-F_{2}^{^3{\rm H}})/x$ is given by the dotted line.
For parton densities in the proton, the CTEQ5L parameterizations are used. 
}
\end{figure}

\begin{figure}
\epsfig{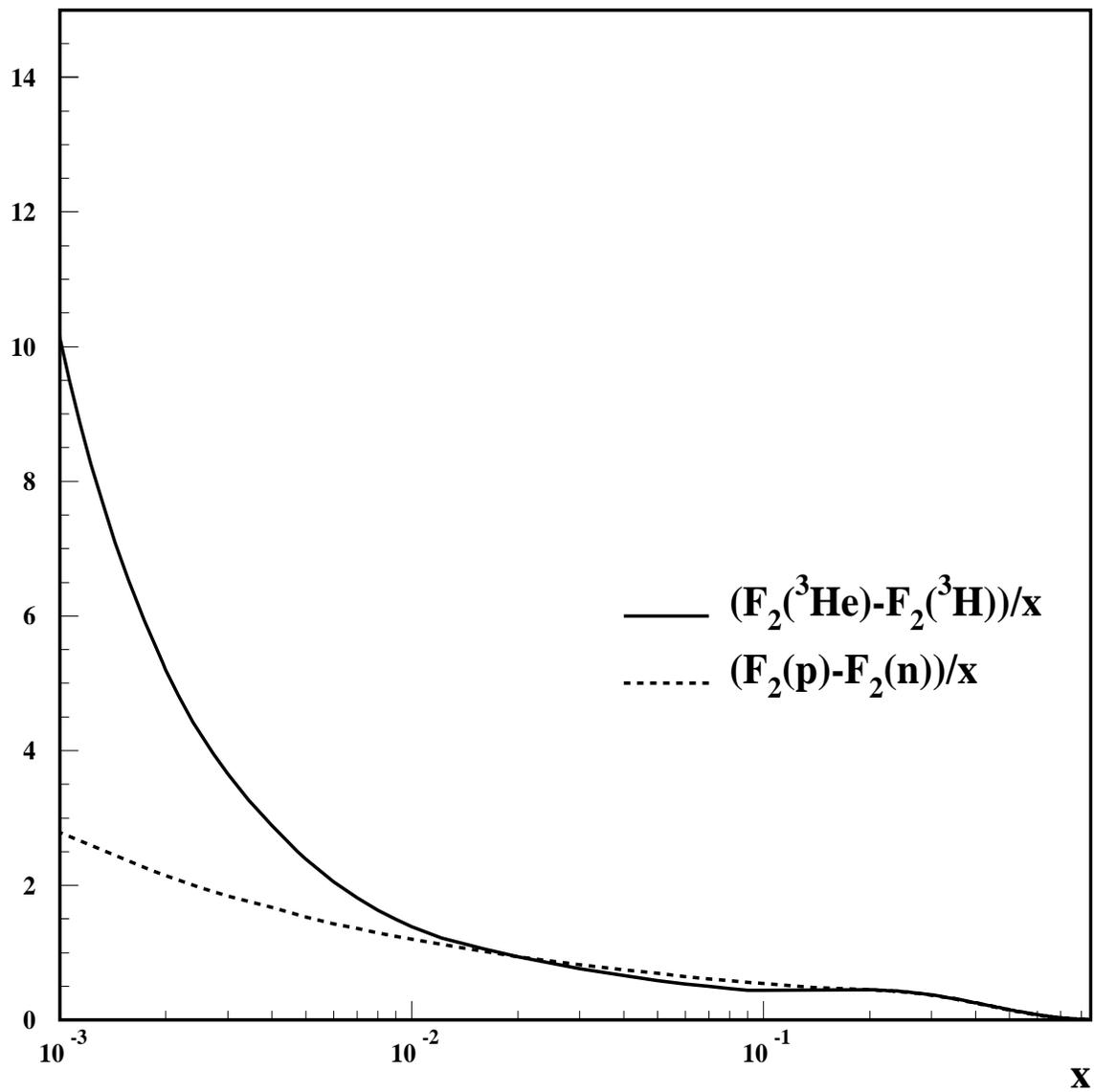}
\label{fig:v2}
\vspace{1cm}
\caption{Case 2. The solid (dashed) line represents $(F_{2}^{^3{\rm He}}-F_{2}^{^3{\rm H}})/x$ ($(F_{2}^{p}-F_{2}^{n})/x$) as a function of Bjorken $x$
at $Q^2$=4 GeV$^2$.
. For parton densities in the proton, the CTEQ5L parameterizations are used.}
\end{figure}

\begin{figure}
\epsfig{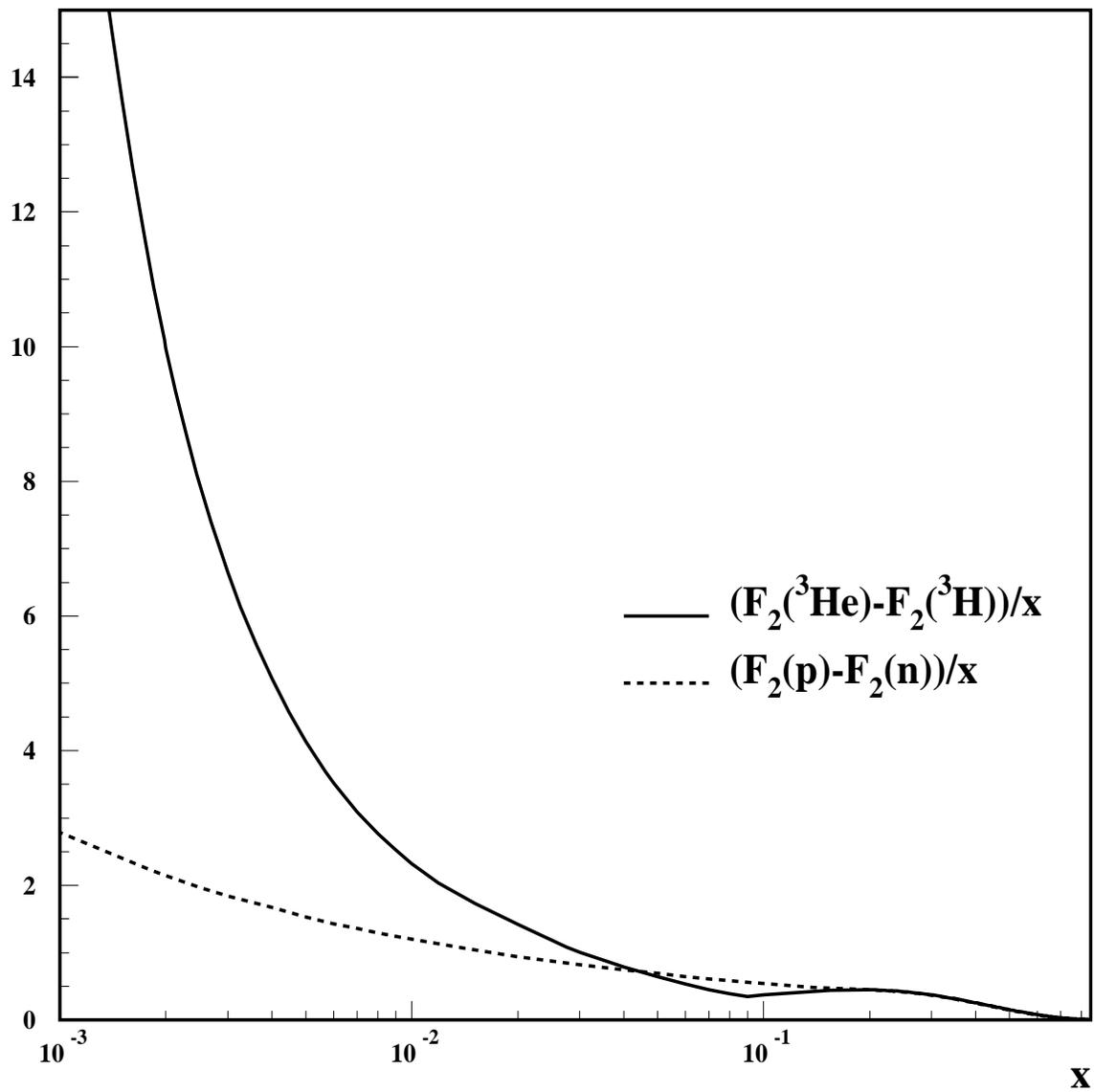}
\label{fig:v3}
\vspace{1cm}
\caption{Case 3. The solid (dashed) line represents $(F_{2}^{^3{\rm He}}-F_{2}^{^3{\rm H}})/x$ ($(F_{2}^{p}-F_{2}^{n})/x$) as a function of Bjorken $x$ at $Q^2$=4 GeV$^2$. For parton densities in the proton, the CTEQ5L parameterizations  are used.}
\end{figure}

\begin{figure}
\epsfig{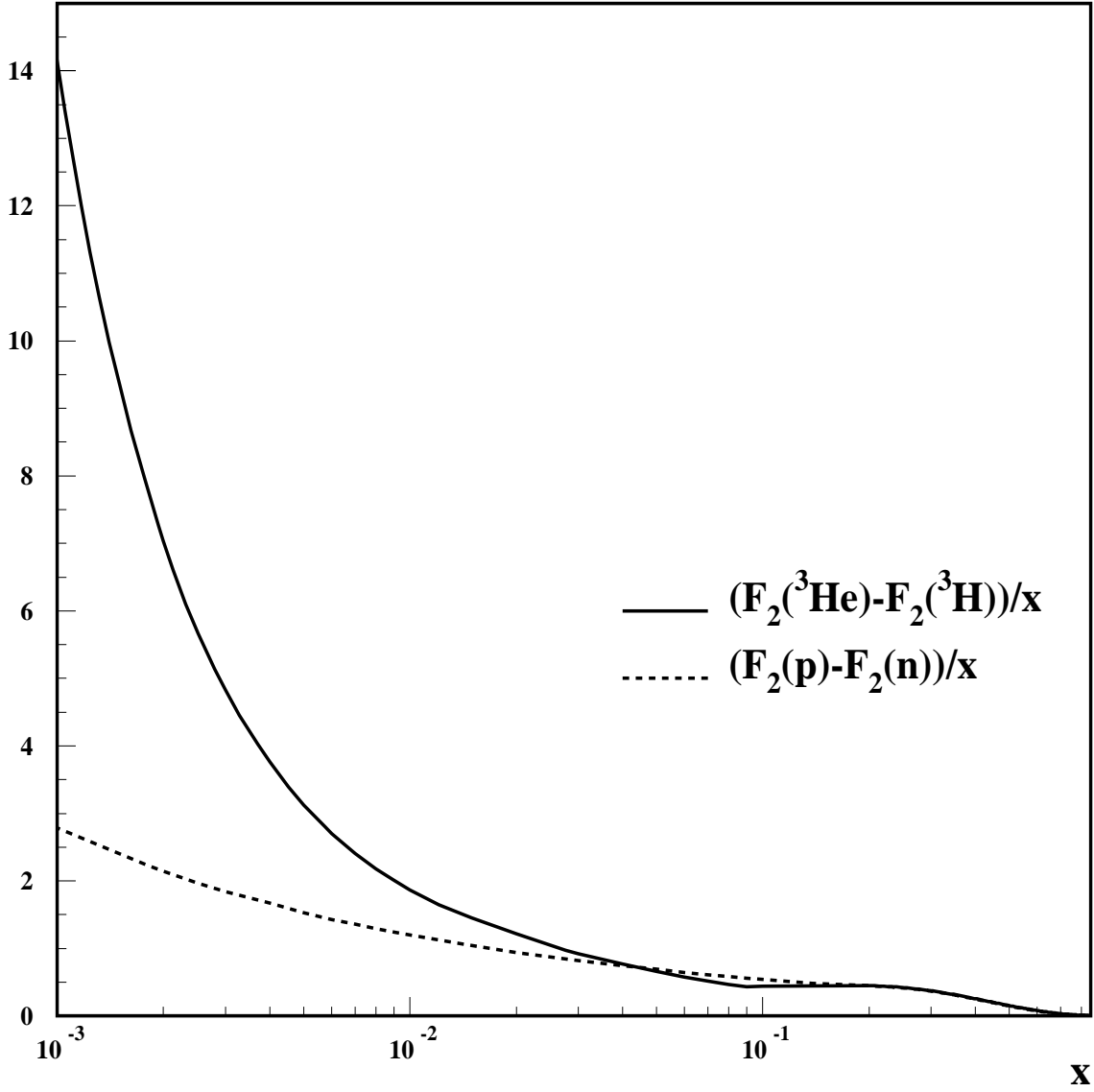}
\label{fig:v4}
\vspace{1cm}
\caption{Case 4. The solid (dashed) line represents $(F_{2}^{^3{\rm He}}-F_{2}^{^3{\rm H}})/x$ ($(F_{2}^{p}-F_{2}^{n})/x$) as a function of Bjorken $x$ at $Q^2$=4 GeV$^2$. For parton densities in the proton, the CTEQ5L parameterizations   are used.}
\end{figure}

\begin{figure}
\epsfig{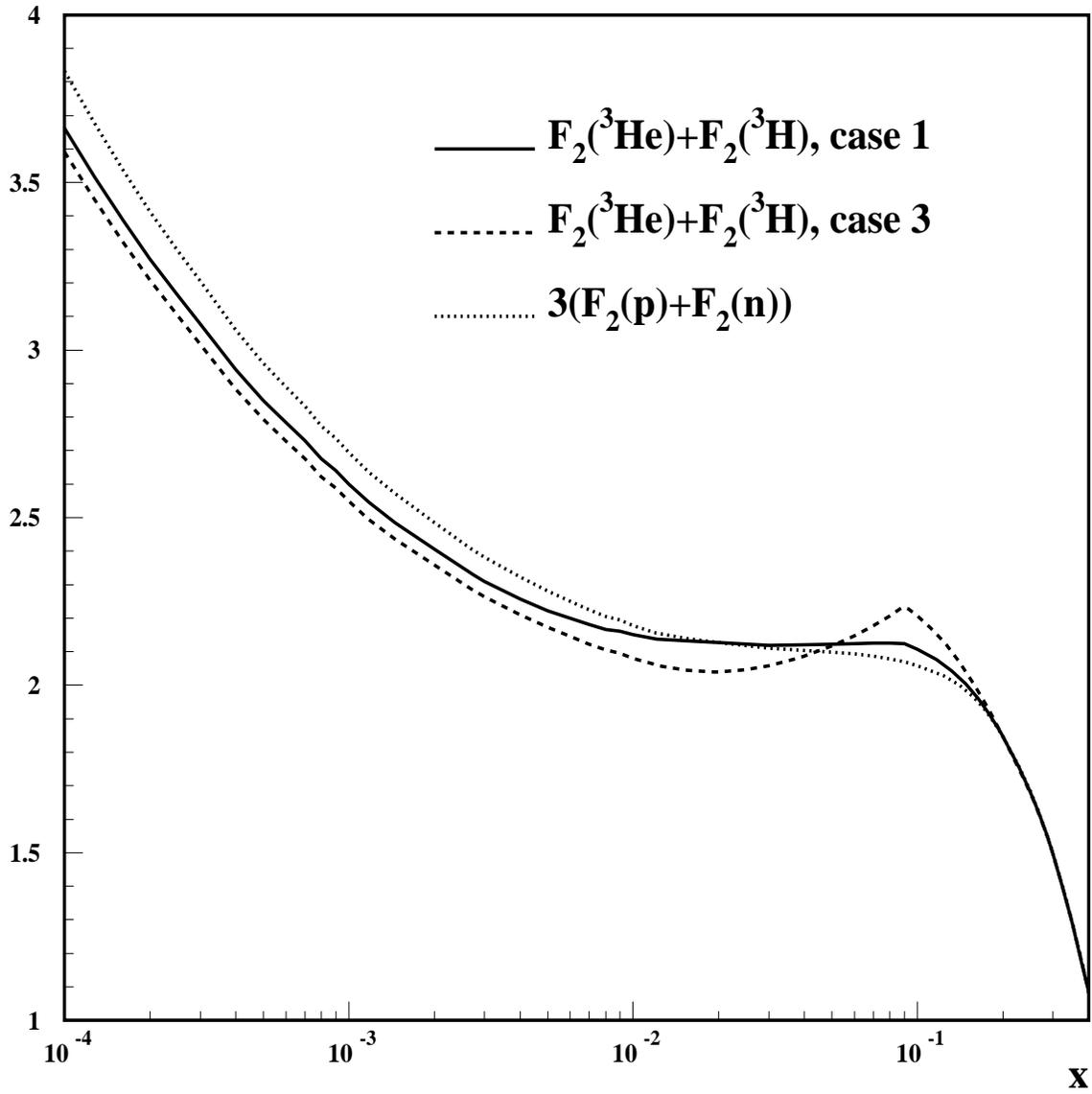}
\label{fig:singlet}
\vspace{1cm}
\caption{The flavor singlet structure functions $F_{2}^{^3{\rm He}}+F_{2}^{^3{\rm H}}$ (solid and dashed lines) and $3(F_{2}^{p}+F_{2}^{n})$ (dotted line) 
as functions of Bjorken $x$ at $Q^2$=4 GeV$^2$.}
\end{figure}


\begin{thebibliography}{99}

\bibitem{NMC1} NMC Coll., P.~Amaudruz {\it et al.},  Phys. Rev. Lett. {\bf 66}~(1991)~2712; 
M.~Arneodo {\it et al.}, Phys. Rev. {\bf D 50}~(1994)~R1.

\bibitem{E866} E866/NuSea Coll., J.C.~Peng {\it et al.}, Phys. Rev. {\bf D 58}~(1998)~092004.

\bibitem{AWT83} A.W.~Thomas, Phys. Lett. {\bf B 126}~(1983)~97.

\bibitem{pion} L.L.~Frankfurt, L.~Mankiewicz, and M.I.~Strikman, Z. Phys. {\bf 334}~(1989)~343; E.M.~Henley and G.A.~Miller, Phys. Lett. {\bf B 251}~(1990)~453; A.I.~Signal, A.W.~Schreiber, and A.W.~Thomas, Mod. Phys. Lett. {\bf A 6}~(1991)~271; 
S.~Kumano, Phys. Rev. {\bf D 43}~(1991)~3067; 
S.~Kumano and J.T.~Londergan, Phys. Rev. {\bf D 44}~(1991)~717; 
W.-Y.P.~Hwang, J.~Speth, and G.E.~Brown, Z. Phys. {\bf A 339}~(1991)~383; 
W.~Koepf, L.L.~Frankfurt, and M.~Strikman, Phys. Rev. {\bf D 53}~(1996)~2586;
J.~Speth and A.W.~Thomas, Adv. Nucl. Phys. {\bf 24}~(1998)~83.


\bibitem{TMS00} A.W.~Thomas, W.~Melnitchouk, and F.M.~Steffens, Phys. Rev. Lett. {\bf 85}~(2000)~2892.

\bibitem{RS79} R.D.~Field and R.P.~Feynman, Phys. Rev. {\bf D 15}~(1977)~2590; 
D.A.~Ross and C.T.~Sachrajda, Nucl. Phys. {\bf B 149}~(1979)~497.

\bibitem{ST89} A.I.~Signal and A.W.~Thomas, Phys. Rev. {\bf D 40}~(1989)~2832.

\bibitem{Kumano} S.~Kumano, Phys. Rep. {\bf 303}~(1998)~183.

\bibitem{Vogt} R.~Vogt, Prog. Part. Nucl. Phys. {\bf 45}~(2000)~S105, also preprint hep-ph/0011298. 

\bibitem{soliton} M.~Wakamatsu, Phys. Rev. {\bf D 44}~(1991)~R2631;
Phys. Lett. {\bf B 269}~(1991)~394; Phys. Rev. {\bf D 46}~(1992)~3762; D.~Diakonov, V.~Petrov, P.Pobylitsa, M.~Polyakov, C.~Weiss, Nucl. Phys. {\bf B 480}~(1996)~341; P.~Pobylitsa, M.~Polyakov, K.~Goeke, T.~Watabe, and C.~Weiss, Phys. Rev. {\bf D 59}~(1999)~034024.

\bibitem{Diakonov} D.I.~Diakonov, V.Yu.~Petrov, and P.V.~Pobylitsa, Nucl. Phys. {\bf B 306}~(1988)~809. 

\bibitem{Bissey} K.~Saito, C.~Boros, K.~Tsushima, F.~Bissey, I.R.~Afnan, and A.W.~Thomas, Phys. Lett. {\bf B 493} (2000)~288. 

\bibitem{FS88} L.~Frankfurt and M.~Strikman, Phys. Rep. {\bf 160}~(1988)~235; Nucl . Phys. {\bf B 316}~(1989)~340.

\bibitem{tjlab} See, for example, ``The Science Driving the 12 GeV 
Upgrade of CEBAF'', edited by the 12 GeV upgrade white paper steering 
committee (TJLAB, 2000). 

\bibitem{RIKEN1} MUSES (Multi-Use Experimental Storage-rings) project at RIKEN, see the RIKEN home page (http://www.rarf.riken.go.jp).

\bibitem{Venugopalan} See, for example, R.~Venugopalan, preprint hep-ph/0102087, and references therein.

\bibitem{PW} G.~Piller and W.~Weise, Phys. Rep. {\bf 330}~(2000)~1;
 D.F.~Geesaman, K.~Saito, and A.W.~Thomas, Ann. Rev. Nucl. Part. Sci. {\bf 45}~(1995)~337; M.~Arneodo, Phys. Rep. {\bf 240}~(1994)~301. 


\bibitem{Gribov} V.N.~Gribov, Sov. J. Nucl. Phys. {\bf 9}~(1969)~369; Sov. Phys. JETP {\bf 29}~(1969)~483; ibid {\bf 30}~(1970)~709.

\bibitem{Glauber} R.J.~Glauber, Phys. Rev. {\bf 100}, 242 (1956).

\bibitem{FGS96} L.~Frankfurt, V.~Guzey, and M.~Strikman, Phys. Lett. {\bf B 381}, 379 (1996).

\bibitem{GS99} V.~Guzey and M.~Strikman, Phys. Rev. {\bf C 61}, 014002, (2000). 

\bibitem{EPW} J.~Edelmann, G.~Piller, and W.~Weise, Z. Phys. {\bf A 357}, 129 (1997); J.~Edelmann, G.~Piller, and W.~Weise, Phys. Rev. {\bf C 57}, 3392 (1998).

\bibitem{FS99} L.~Frankfurt and M.~Strikman, Eur. Phys. J. {\bf A 5}~(1999)~293.

\bibitem{MT} W.~Melnitchouk and A.W.~Thomas, Phys. Lett. {\bf B 317}~(1993)~437; W.~Melnitchouk and A.W.~Thomas, Phys. Rev. {\bf C 52}~(1995)~3373; J.~Kwiecinski and B.~Badelek, Phys. Lett. {\bf B 208}, 508 (1988). 

\bibitem{NMC} NMC Collab., P.~Amaudruz {\it et al.}, Nucl. Phys. {\bf B 441}, 3 (1995).


\bibitem{LS76} E.~Levin and M.~Strikman, Sov. J. Nucl. Phys. {\bf 23}, 216 (1976).

\bibitem{BJ77} R.C.~Barrett and D.F.~Jackson {\it Nuclear sizes and structure}, Clarendon Press, Oxford, 1977.

\bibitem{Rosenfelder} R.~Rosenfelder, Phys. Lett. {\bf B 479}~(2000)~381.

\bibitem{AD} H.~De Vries, C.W~De Jager, and C.~De Vries, Atomic Data and Nuclear Data Tables {\bf 36}~(1987)~495.

\bibitem{FLS} L.L.~Frankfurt, M.I.~Strikman, and S.~Liuti, Phys. Rev. Lett. {\bf 65}~(1990)~1725.

\bibitem{Eskola} K.J.~Eskola, V.J.~Kolhinen, and C.A.~Saldago, Eur. Phys. J. {\bf C 9}~(1999)~61; K.J.~Eskola, V.J.~Kolhinen, and P.V.~Ruuskanen, Nucl. Phys. {\bf B 535}~(1998)~351.

\bibitem{CTEQ} H.L.~Lai {\it et al.}, Eur. Phys. J. {\bf C 12}~(2000)~375.

\bibitem{MFGS} M.~McDermott, L.~Frankfurt, V.~Guzey, M.~Strikman, Eur. Phys. J. {\bf C 16}~(2000)~641.  

\bibitem{HP} B.~Povh and J.~H\"{u}fner, Phys. Rev. Lett. {\bf 58}~(1987)~1612.
 

\end{thebibliography}
\end{document}